\documentclass{article}
\usepackage[T1]{fontenc} 
\usepackage[utf8]{inputenc} 
\usepackage[lbd]{ismir}
\usepackage{amsmath,cite,url}
\usepackage{graphicx}
\usepackage{color}
\usepackage{listings}
\usepackage{xcolor}
\usepackage{comment}
\definecolor{cmtgray}{gray}{0.42}
\lstdefinestyle{layerit}{%
  language=Python,
  basicstyle=\fontfamily{zi4}\fontsize{8}{9.4}\selectfont,
  commentstyle=\color{cmtgray}\itshape,
  keywordstyle=\bfseries,
  showstringspaces=false,
  columns=fullflexible, keepspaces=true,
  breaklines=false,                     
  aboveskip=0pt, belowskip=0pt, xleftmargin=0pt, numbers=none}
\usepackage{musicography}

\lstdefinestyle{layeritinline}{style=layerit,
   basicstyle=\ttfamily\scriptsize,
   aboveskip=6pt, belowskip=4pt, xleftmargin=12pt}

\title{LayerIt: Towards a Framework for Time-Aligned, Composable Music Visualizations}

\oneauthor
  {Fernando Azeredo \hspace{1cm} Ant\'onio S\'a Pinto}
  {Faculdade de Engenharia da Universidade do Porto, Portugal \\
   INESC TEC, Porto, Portugal \\
   {\tt up202308655@fe.up.pt}}

\def\authorname{F. Azeredo and A. S\'a Pinto}

\usepackage[bookmarks=false,pdfauthor={\authorname},hidelinks,pdftitle={LayerIt: Towards a Framework for Time-Aligned, Composable Music Visualizations}]{hyperref}

\begin{document}

\maketitle
\begin{abstract}
Music information retrieval often relates signal-derived, algorithmic, and symbolic information across different coordinate systems. Existing visualizations typically leave notation separate from physical time, while composites that combine them are assembled by hand. We present LayerIt, an open-source Python library for composing independent representations with notation on a shared performance-time axis. Given a score warped into performance time and a note-level alignment, LayerIt emits a single SVG that preserves the score’s MEI structure and keeps added components identifiable and restylable. We demonstrate this through a challenging beat-tracking analysis, in which tracker output, signal representations, and notation can be inspected together to locate metrical disagreement and other errors against both notated structure and performed time.
\end{abstract}

\section{Introduction}\label{sec:introduction}

Music can be represented in many ways: as engraved notation, symbolic encodings, or audio recordings of performances. These representations, however, live in different notions of time. Recordings unfold in physical time, measured in seconds or samples; symbolic encodings organize it in abstract units such as beats and measures; and engraved notation renders it in space, in pixels or millimetres. Establishing correspondences between them, a task known as alignment or synchronization, is central to both music information retrieval (MIR) and digital musicology.

Alignment algorithms, notably those matching audio to score, must balance accuracy against computational cost and robustness \cite{muller2021SyncToolboxPython}. The latter takes several forms, such as tolerance to performance errors \cite{nakamura2017PerformanceErrorDetection} or to structural mismatches between score and performance \cite{fremerey2010HandlingRepeatsJumps}. Two further distinctions cut across that balance: matching may happen within the audio or symbolic domain, or across the two \cite{mueller2019CrossModalMusicRetrieval}, and may rely on classical signal-processing features or on learned representations \cite{zeitler2025RobustAccurateAudio}.

An alignment, once computed, can be put to work in more than one way. One concern is representation. Alignments are often stored in ad hoc, single-use formats, and recent modelling relates musical timelines across the graphical, logical, and physical domains so that positions and annotations move between them and the alignment can be reused \cite{hentschel2026TimeAlignModelling}. A separate concern, which this work takes up, is visualization. The prevailing strategy fixes the score's layout and links its elements to instants on a physical timeline through drawn connectors \cite{thomas2012LinkingSheetMusic}, as do interactive tools that track a performance with a moving score cursor \cite{jiang2025PianoPrecisionAdvancing}. Score and physical time thus stay in separate coordinate systems, leaving the user to bridge them mentally.

An alternative dispenses with anchors altogether. The notation itself is laid out in performance time, each element repositioned according to its performed onset, so that the score stays fully readable as notation while doubling as a visual index into physical time \cite{goebl2025LetsScoreWarpAgain}. The possibility that motivates the present work is a shared performance-time axis onto which further material can be layered, from waveforms and feature curves to other time-aligned information~\cite{pinto2020ShiftIfYou}. Annotations such as beat positions and metre are of particular interest here, since they are explicit in musical notation yet recoverable from audio only by estimation.

No simple, programmatic way to compose these layers reproducibly exists. Such composites are assembled by hand, score and audio matched by eye and ear, so nothing ties the match to the signal and nothing survives a change of recording or algorithm.
We present LayerIt\footnote{\url{https://github.com/FernandoAze/LayerIt}}: given a score--performance alignment, it places independently generated components onto this shared axis and merges them into a single, self-contained visualization. Each is handled as an independent layer that can be added, combined, or removed, making such composites straightforward to build and adapt. The library is open-source and composes the figure in Python, where the material is already being produced, so that no export to a separate drawing tool is required.

\section{The LayerIt Framework}\label{sec:framework}
A LayerIt figure is a stack of panels that share one performance-time axis.
Four inputs define it: a performance recording, read with librosa \cite{mcfee2015LibrosaAudioMusic}; a score already warped onto that axis by ScoreWarp \cite{goebl2025LetsScoreWarpAgain}; a note-level alignment of onsets to score elements \cite{weigl2020MultimodalMusicInformation}, which LayerIt consumes rather than computes; and additional data representations.
The output is a single SVG document.

Two numbers from the score's time axis, its span in seconds and its length in pixels, yield the system's only scale factor. Nothing else is warped: each layer converts its native units (e.g.\ samples, frames) into seconds where needed, and is drawn at that scale. A layer therefore sets neither placement; the horizontal follows from the scale, the vertical from the composition.

A representation is a class with two methods: one loads and validates its data, the other draws it into a Matplotlib\,\texttt{Axes}~\cite{hunter2007Matplotlib2DGraphics}, so that existing plotting knowledge transfers unchanged. Placed in a panel it becomes a layer, and an optional third method lets it emit its own SVG group. Ten representations are implemented on this interface, based on four shapes, an initial, extensible set documented in the repository: \texttt{Curve} (any function of time), \texttt{Events} (instants), \texttt{Intervals} (time spans), and \texttt{Field} (a dense time--frequency/pitch array). No representation touches the aligning machinery.

The score is copied into the composite rather than redrawn from its Verovio rendering \cite{pugin2014VerovioLibraryEngraving}, thereby preserving its Music Encoding Initiative (MEI) identifiers and hierarchy.
Each added component is emitted as its own named group, with identifiers rewritten per panel so that repeated layers stay distinguishable; the vector shapes remain native SVG primitives, selectable and restylable, while a field embeds a raster image inside its group.

\section{Demonstration}\label{sec:demonstration}
We demonstrate LayerIt on the figure a beat-tracking analysis calls for: a recording, a score, and a tracker's frame-level output.
The example uses the first six bars of Debussy's \emph{Clair de Lune} performed by Maria João Pires, the MEI score to which its onsets were aligned in Piano Precision~\cite{jiang2025PianoPrecisionAdvancing}, and the beat and downbeat activations and estimates from Beat This!~\cite{foscarin2024BeatThisAccurate}.

\begin{figure}
\includegraphics[alt={Four panels share one horizontal time axis of about 33
seconds. Top: the opening six bars of the piano score in 9/8, marked Andante
très expressif, thus spaced unevenly in time (as played). Second: two activation curves, mostly low with sharp peaks, crossed by
dashed vertical lines at the notated onsets. Third: the audio waveform overlaid
with a sparse set of downbeat markers and a much denser set of beat markers,
several falling within a single notated beat. Bottom: a mel spectrogram from
100 to 1500 Hz. The beat estimates greatly outnumber the beats notated above
them, and many are incorrectly identified as downbeats.},width=.9\columnwidth]{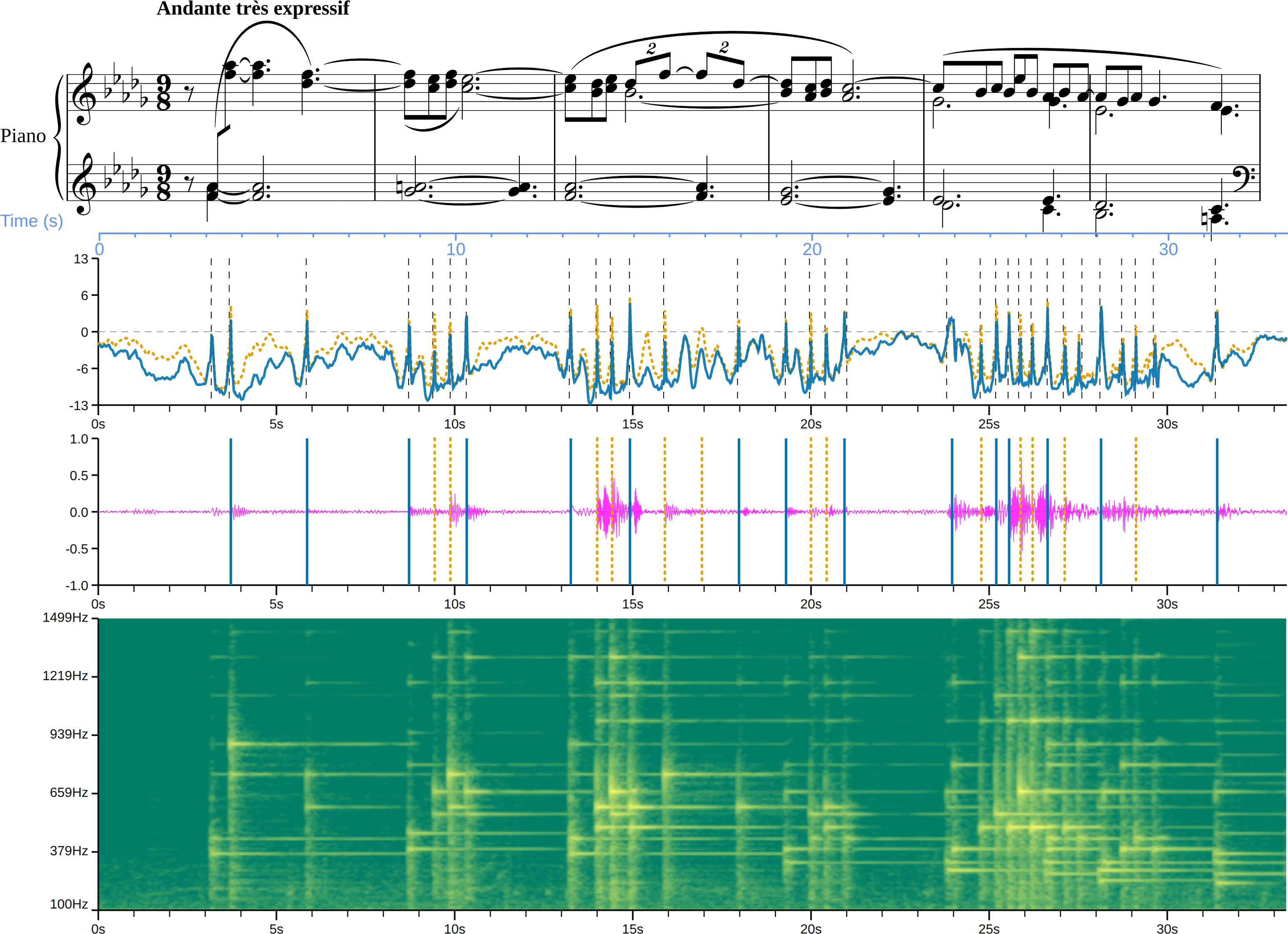}
\caption{LayerIt composition, top to bottom: 1) warped score;
2) beat/downbeat logits with score onsets dashed;
3) waveform in pink with beat/downbeat estimates;
4) mel spectrogram. Beats dotted orange, downbeats solid blue.}
\label{fig:composite}
\end{figure}

The abridged script (with styling arguments omitted) that produces \figref{fig:composite} requires only five statements:
\begin{lstlisting}[style=layeritinline]
fig = Visualizer(audio, score, maps, beats)
fig.add_panel(Onset(), BeatLogits(), 
    DownbeatLogits(), height_scale=.5)
fig.add_panel(Waveform(), BeatsLayer(), 
    DownbeatsLayer(), height_scale=.5)
fig.add_panel(MelSpec(freq_window=(100, 1500)))
fig.compose("LBD_FIG.svg")
\end{lstlisting}

Most downbeats are correctly placed, but the many false positives---several of them quaver onsets taken as beats---show that the metrical structure is misread.
Two conditions known to trouble beat trackers hold here: compound triple metre (given its scarcity in training datasets) and highly expressive interpretations~\cite{pinto2021TappingDifficultOnes}.
Both are apparent in the figure: compound triple metre in its 9/8 time signature, expressive playing in its \emph{très expressif} marking. The performer stretches and compresses the beat~(\musQuarter.), as the warped score's spacing shows.
The estimates, read against the notation, become an account of the difficulties behind them, not a list of errors. Each false positive carries a musical address, the bar it falls in and the figure that produced it, so the passage can be found in the score and heard in the recording. A signal-derived view shows only where they fall.

\section{Conclusion}\label{sec:conclusion}
LayerIt replaces the hand assembly of score-aligned MIR figures with a programmatic specification. Beat-tracking output is ordinarily examined against the signal (e.g.\ in Sonic Visualiser~\cite{cannam2010SonicVisualiserOpen}), while the score, if consulted, is read separately and the correspondence reconstructed by the user.
LayerIt instead puts both on the same performance-time axis, where the tracker’s inferred metrical structure can be read against the score’s, and seen to diverge. Neither carries what the other does. Beat positions and their metrical organization are notated but only inferable from the signal; performed timing is in the recording alone. The comparison needs both, aligned. Beat tracking is one instance of a broader case: any analysis that must relate algorithmic or signal-derived output to notated structure in performance time.

Two limitations bear on use. The framework depends on an MEI score and an externally computed alignment, inheriting errors in both, and the warped layout may carry artefacts absent from conventional engraving. More narrowly, dense time--frequency/pitch fields are rasterized and cannot be restyled like the vector components.

The most immediate of the open directions concerns alignment input. LayerIt reads the onset-level format consumed by ScoreWarp; integration with the Sync Toolbox~\cite{muller2021SyncToolboxPython} or Time to Align!~\cite{hentschel2026TimeAlignModelling} would do more than widen it. The former could add automatic synchronization, while the latter could retain provenance and domain information absent from an onset list. A further direction concerns annotation: aligned symbolic structure and signal-based evidence may support interactive workflows that a divided view makes difficult.

\newpage

\bibliography{ISMIRtemplate_ASP_v2}

@InProceedings{	  cannam2010SonicVisualiserOpen,
  title		= {{Sonic Visualiser: An Open Source Application for Viewing,
		  Analysing, and Annotating Music Audio Files}},
  booktitle	= {Proceedings of the 18th {{ACM International Conference}}
		  on {{Multimedia}} ({{MM}} '10)},
  author	= {Cannam, Chris and Landone, Christian and Sandler, Mark},
  year		= 2010,
  pages		= {1467--1468},
  publisher	= {ACM Press},
  address	= {Florence, Italy},
  doi		= {10.1145/1873951.1874248}
}

@InProceedings{	  foscarin2024BeatThisAccurate,
  title		= {{Beat This! Accurate Beat Tracking without DBN
		  Postprocessing}},
  booktitle	= {Proceedings of the 25th {{International Society}} for
		  {{Music Information Retrieval Conference}} ({{ISMIR}})},
  author	= {Foscarin, Francesco and Schl{\"u}ter, Jan and Widmer,
		  Gerhard},
  year		= 2024,
  primaryclass	= {cs},
  address	= {San Francisco, CA, USA},
  doi		= {10.48550/arXiv.2407.21658},
  urldate	= {2025-02-04},
  archiveprefix	= {arXiv}
}

@InProceedings{	  fremerey2010HandlingRepeatsJumps,
  title		= {Handling {{Repeats}} and {{Jumps}} in {{Score-Performance
		  Synchronization}}},
  booktitle	= {Proceedings of the 11th {{International Society}} for
		  {{Music Information Retrieval Conference}} ({{ISMIR}})},
  author	= {Fremerey, Christian and M{\"u}ller, Meinard and Clausen,
		  Michael},
  year		= 2010,
  pages		= {243--248},
  address	= {Utrecht, The Netherlands},
  langid	= {english}
}

@InProceedings{	  goebl2025LetsScoreWarpAgain,
  title		= {{Let's Do the ScoreWarp Again! Shifting Notes to
		  Performance Timelines}},
  booktitle	= {Proceedings of the {{Music Encoding Conference}}},
  author	= {Goebl, Werner and Weigl, David M. and Lewis, David and
		  Plaksin, Anna and Stremel, Sophie},
  year		= 2025,
  address	= {London, United Kingdom},
  doi		= {10.17613/gqd3b-w7c28},
  urldate	= {2026-06-05},
  langid	= {english}
}

@Article{	  hentschel2026TimeAlignModelling,
  title		= {Time to {{Align}}! {{Modelling Musical Timelines}} for
		  {{Music Information Retrieval}} and {{Digital
		  Musicology}}},
  author	= {Hentschel, Johannes and Berndt, Axel and
		  {Cancino-Chac{\'o}n}, Carlos and Dixon, Simon and Foo, Anne
		  and Gotham, Mark and Hu, Patricia and K{\"o}ster, Maik and
		  Martins, Felipe D. and Mauro, Davide A. and M{\"u}ller,
		  Meinard and Neuwirth, Markus and Pacha, Alexander and Page,
		  Kevin R. and Peter, Silvan and Polyakov, Egor and Pugin,
		  Laurent and Weigl, David M. and Wei{\ss}, Christof and
		  Widmer, Gerhard},
  year		= 2026,
  journal	= {Transactions of the International Society for Music
		  Information Retrieval},
  volume	= {9},
  number	= {1},
  pages		= {384--404},
  doi		= {10.5334/tismir.296},
  urldate	= {2026-08-19},
  langid	= {english}
}

@Article{	  hunter2007Matplotlib2DGraphics,
  title		= {Matplotlib: {{A 2D Graphics Environment}}},
  shorttitle	= {Matplotlib},
  author	= {Hunter, John D.},
  year		= 2007,
  journal	= {Computing in Science \& Engineering},
  volume	= {9},
  number	= {3},
  pages		= {90--95},
  doi		= {10.1109/MCSE.2007.55},
  urldate	= {2026-08-19},
  copyright	= {https://ieeexplore.ieee.org/Xplorehelp/downloads/license-information/IEEE.html}
}

@InProceedings{	  jiang2025PianoPrecisionAdvancing,
  title		= {Piano {{Precision}}: {{Advancing Music Performance
		  Analysis}} by {{Integrating User-Correctable Audio-to-Score
		  Alignment}}},
  shorttitle	= {Piano {{Precision}}},
  booktitle	= {Proceedings of the 22nd {{Sound}} and {{Music Computing
		  Conference}} ({{SMC}})},
  author	= {Jiang, Yucong and Weigl, David M. and Goebl, Werner},
  year		= 2025,
  pages		= {160--167},
  publisher	= {{Institute of Electronic Music and Acoustics, University
		  of Music and Performing Arts Graz}},
  address	= {Graz, Austria},
  doi		= {10.5281/ZENODO.15838699},
  urldate	= {2026-06-05},
  copyright	= {Creative Commons Attribution 4.0 International},
  langid	= {english}
}

@InProceedings{	  mcfee2015LibrosaAudioMusic,
  title		= {Librosa: {{Audio}} and {{Music Signal Analysis}} in
		  {{Python}}},
  booktitle	= {Proceedings of the 14th {{Python}} in {{Science
		  Conference}}},
  author	= {McFee, Brian and Raffel, Colin and Liang, Dawen and Ellis,
		  Daniel P W and McVicar, Matt and Battenberg, Eric and
		  Nieto, Oriol},
  year		= 2015,
  pages		= {18--25},
  address	= {Austin, Texas, USA}
}

@Article{	  mueller2019CrossModalMusicRetrieval,
  title		= {Cross-{{Modal Music Retrieval}} and {{Applications}}: {{An
		  Overview}} of {{Key Methodologies}}},
  shorttitle	= {Cross-{{Modal Music Retrieval}} and {{Applications}}},
  author	= {Mueller, Meinard and Arzt, Andreas and Balke, Stefan and
		  Dorfer, Matthias and Widmer, Gerhard},
  year		= 2019,
  journal	= {IEEE Signal Processing Magazine},
  volume	= {36},
  number	= {1},
  pages		= {52--62},
  doi		= {10.1109/MSP.2018.2868887},
  urldate	= {2026-08-19}
}

@Article{	  muller2021SyncToolboxPython,
  title		= {Sync {{Toolbox}}: {{A Python Package}} for {{Efficient}},
		  {{Robust}}, and {{Accurate Music Synchronization}}},
  shorttitle	= {Sync {{Toolbox}}},
  author	= {M{\"u}ller, Meinard and {\"O}zer, Yigitcan and Krause,
		  Michael and Pr{\"a}tzlich, Thomas and Driedger, Jonathan},
  year		= 2021,
  journal	= {Journal of Open Source Software},
  volume	= {6},
  number	= {64},
  pages		= {3434},
  doi		= {10.21105/joss.03434},
  urldate	= {2026-08-19},
  langid	= {english}
}

@InProceedings{	  nakamura2017PerformanceErrorDetection,
  title		= {Performance {{Error Detection}} and {{Post-Processing}}
		  for {{Fast}} and {{Accurate Symbolic Music Alignment}}},
  booktitle	= {Proceedings of the 18th {{International Society}} for
		  {{Music Information Retrieval Conference}} ({{ISMIR}})},
  author	= {Nakamura, Eita and Yoshii, Kazuyoshi and Katayose,
		  Haruhiro},
  year		= 2017,
  pages		= {347--353},
  address	= {Suzhou, China},
  urldate	= {2026-08-19}
}

@InProceedings{	  pinto2020ShiftIfYou,
  title		= {Shift {{If You Can}}: {{Counting}} and {{Visualising
		  Correction Operations}} for {{Beat Tracking Evaluation}}},
  booktitle	= {Extended {{Abstracts}} for the {{Late-Breaking Demo
		  Session}} of the 21st {{International Society}} for {{Music
		  Information Retrieval Conference}} ({{ISMIR}})},
  author	= {Pinto, Ant{\'o}nio S{\'a} and Domingues, In{\^e}s and
		  Davies, Matthew E. P.},
  year		= 2020,
  doi		= {10.48550/arXiv.2011.01637}
}

@InCollection{	  pinto2021TappingDifficultOnes,
  title		= {Tapping {{Along}} to the {{Difficult Ones}}: {{Leveraging
		  User-Input}} for {{Beat Tracking}} in {{Highly Expressive
		  Musical Content}}},
  booktitle	= {Perception, {{Representations}}, {{Image}}, {{Sound}},
		  {{Music}} -- 14th {{International Symposium}}, {{CMMR}} 2019,
		  {{Marseille}}, {{France}}, {{October}} 14--18, 2019, {{Revised
		  Selected Papers}}},
  author	= {Pinto, Ant{\'o}nio S{\'a} and Davies, Matthew E. P.},
  editor	= {{Kronland-Martinet}, Richard and Ystad, S{\o}lvi and
		  Aramaki, Mitsuko},
  year		= 2021,
  series	= {Lecture {{Notes}} in {{Computer Science}}},
  volume	= {12631},
  pages		= {75--90},
  publisher	= {Springer},
  doi		= {10.1007/978-3-030-70210-6_5},
  isbn		= {978-3-030-70209-0}
}

@InProceedings{	  pugin2014VerovioLibraryEngraving,
  title		= {Verovio: {{A Library}} for {{Engraving MEI Music
		  Notation}} into {{SVG}}},
  booktitle	= {Proceedings of the 15th {{International Society}} for
		  {{Music Information Retrieval Conference}} ({{ISMIR}})},
  author	= {Pugin, Laurent and Zitellini, Rodolfo and Roland, Perry},
  year		= 2014,
  pages		= {107--112},
  address	= {Taipei, Taiwan},
  langid	= {english}
}

@InCollection{	  thomas2012LinkingSheetMusic,
  title		= {{Linking Sheet Music and Audio -- Challenges and New
		  Approaches}},
  booktitle	= {Multimodal {{Music Processing}}},
  author	= {Thomas, Verena and Fremerey, Christian and M{\"u}ller,
		  Meinard and Clausen, Michael},
  editor	= {M{\"u}ller, Meinard and Goto, Masataka and Schedl, Markus},
  year		= 2012,
  series	= {Dagstuhl {{Follow-Ups}}},
  volume	= {3},
  pages		= {1--22},
  publisher	= {Schloss Dagstuhl -- Leibniz-Zentrum f\"ur Informatik},
  address	= {Dagstuhl, Germany},
  doi		= {10.4230/DFU.Vol3.11041.1},
  urldate	= {2026-08-19}
}

@TechReport{	  weigl2020MultimodalMusicInformation,
  title		= {Multimodal {{Music Information Alignment}} {{v2}}},
  author	= {Weigl, David M.},
  year		= 2020,
  number	= {Deliverable 3.5},
  institution	= {TROMPA}
}

@InProceedings{	  zeitler2025RobustAccurateAudio,
  title		= {Robust and {{Accurate Audio Synchronization Using Raw
		  Features}} from {{Transcription Models}}},
  booktitle	= {Proceedings of the 26th {{International Society}} for
		  {{Music Information Retrieval Conference}} ({{ISMIR}})},
  author	= {Zeitler, Johannes and Maman, Ben and M{\"u}ller, Meinard},
  year		= 2025,
  pages		= {120--127},
  address	= {Daejeon, South Korea},
  langid	= {english}
}

\end{document}